\documentclass[12pt]{article}
\usepackage{pst-plot,url,epsf}
\newcommand{\beq}{\begin{equation}}
\newcommand{\eeq}{\end{equation}}
\newcommand{\bea}{\begin{eqnarray}}
\newcommand{\eea}{\end{eqnarray}}
\newcommand{\beas}{\begin{eqnarray*}}
\newcommand{\eeas}{\end{eqnarray*}}

\newcommand{\epm}{e^+e^-}

\newcommand{\ra}{\rightarrow}

\newcommand{\AmS}{{\protect\the\textfont2
  A\kern-.1667em\lower.5ex\hbox{M}\kern-.125emS}}
\newcommand{\nn}{\nonumber}

\begin{document}
\thispagestyle{empty}
\hspace*{15.cm} March 2008\\
\hspace*{15.cm} Revised:\\
\hspace*{15.cm} April 2008\\[1.5cm]
\begin{center}
{\LARGE\bf Off resonance background effects in $\epm \ra t \bar t H$}\\
\vspace*{2cm}
%----------------------------
Karol Ko\l odziej\footnote{E-mail: karol.kolodziej@us.edu.pl} 
and Szymon Szczypi\'nski\footnote{E-mail: simon@server.phys.us.edu.pl}\\[6mm]
{\small\it Institute of Physics, University of Silesia\\ 
ul. Uniwersytecka 4, PL-40007 Katowice, Poland}\\
\vspace*{2.5cm}
{\bf Abstract}\\
\end{center}
The top quark Yukawa coupling to the intermediate mass Higgs boson
can be determined in the reaction $\epm \ra t\bar t H$ that,
after taking into account decays, will be
detected at the International Linear Collider through reactions with 8 particles 
in the final state. Such $2 \ra 8$ reactions receive contributions 
from tens thousands of Feynman diagrams,
already in the lowest order of Standard Model,  most of which comprise background
to resonant associated production and decay of the top quark pair and Higgs boson.
We illustrate the background effects by comparing cross sections 
of three reactions, which represent different detection channels
of $\epm \ra t\bar t H$, calculated with the complete sets of the lowest
order Feynman diagrams with the corresponding signal cross 
sections calculated with the diagrams
of associated production and decay of off mass shell top quark pair and Higgs 
boson only. The comparison that is performed with different selections of cuts
shows that the background effects are sizeable, but they can be reduced by appropriate 
choice of cuts.

\vfill

\newpage

\section{Introduction}

\begin{figure}[htb]
\vspace{140pt}
\centerline{
\includegraphics{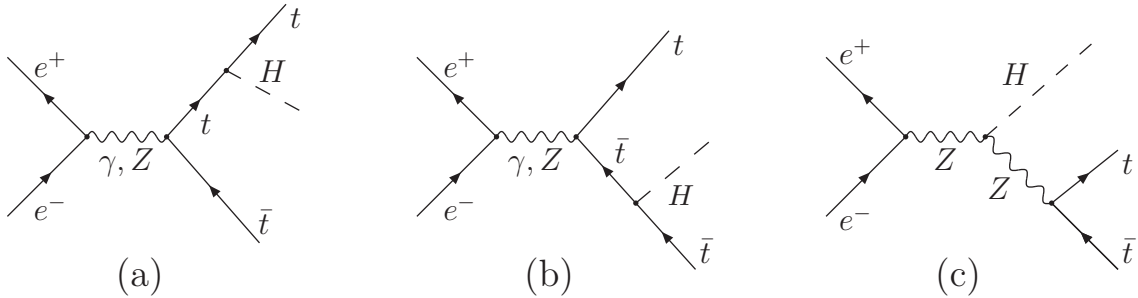}\hfill}
\caption{Feynman diagrams of reaction (\ref{eetth}) to the lowest order of SM
         with the neglect of the Higgs boson coupling to electrons.}
\label{fig:eetth}
\end{figure}

If the Standard Model (SM) Higgs boson has mass below the $t \bar t$ 
threshold, $m_{H} < 2 m_{t}$, then its Yukawa coupling to the top quark 
\bea
g_{ttH}=\frac{m_t}{v}, \qquad {\rm with} \qquad 
v=(\sqrt{2}G_F)^{-1/2}\simeq 246 \rm{GeV},\nn
\eea
which is by far the largest Yukawa coupling of SM, can
be best determined at the future International Linear Collider (ILC) 
\cite{ILC} through measurement of 
the reaction of associated production of the top quark 
pair and Higgs boson \cite{eetth}
\begin{equation}
\label{eetth}
 \epm \ra t \bar{t} H.
\end{equation}
The lowest order SM Feynman diagrams of reaction (\ref{eetth}), with the neglect
of the Higgs boson coupling to electrons, are shown in Fig.~\ref{fig:eetth}.
As the contribution of the Higgs boson emission off the virtual $Z$-boson line, 
which is represented by the diagram in Fig.~\ref{fig:eetth}c,
is small with respect to the Higgsstrahlung off the top quark line illustrated
in Fig.~\ref{fig:eetth}a and \ref{fig:eetth}b, the SM lowest
order cross section of reaction (\ref{eetth}) becomes practically
proportional to $g_{ttH}^2$. This fact makes  
reaction (\ref{eetth}) so sensitive to the top--Higgs Yukawa coupling. 

Particles on the right hand side of reaction (\ref{eetth}) are unstable:
the top and antitop decay, even before they hadronize, predominantly into 
$b W^+$ and $\bar{b} W^-$, respectively, and the Higgs boson, 
dependent on its mass, 
decays either into a fermion--antifermion or an electroweak (EW) gauge 
boson pair
and the EW  bosons subsequently decay, each into a fermion--antifermion pair.
This, dependent on the Higgs boson mass $m_H$, leads to reactions with either 8 or  
10 fermions in the final state. As direct searches for the Higgs boson at LEP, which 
give a lower limit for $m_H$ of $114.4$~GeV at 95\% CL \cite{LEPdir},
combined with theoretical constrains in the framework of SM 
%from the virtual effects it has on precision EW observables:
%the value of the top quark mass measured at Tevatron and the 
%combined $W$-boson mass \cite{Hmass} give a central value
%of $m_H=85_{-28}^{+39}$ GeV and an upper limit of 166~GeV, both at
%95\% CL \cite{Kilminster}. 
favour a value of $m_H$ in the range just above the lower direct search limit
\cite{Hmass}, we will assume  $m_H < 140$~GeV.
Then the Higgs boson would decay dominantly into a $b \bar b$-quark pair
and reaction (\ref{eetth})
will be actually detected at the ILC through reactions of the form
\beq
\label{ee8f}
  e^+e^-\;\; \ra \;\;  b \bar b  b \bar b f_1\bar{f'_1} f_2 \bar{f'_2},
\eeq
where $f_1, f'_2 =\nu_{e}, \nu_{\mu}, \nu_{\tau}, u, c$ and 
$f'_1, f_2 = e^-, \mu^-, \tau^-, d, s$. Thus, reaction (\ref{ee8f}) can
be detected in any of the following channels: the hadronic (38\%),
leptonic (25\%), or  
semileptonic (37\%), corresponding to
decay modes of the $W$-bosons coming from 
decays of the $t$- and $\bar t$-quark of reaction (\ref{eetth}). 
%The semileptonic channel should give the most
%distinct signature of the associated top quark pair Higgs boson production.

Since the original work of the early 1990's \cite{eetth} 
reaction (\ref{eetth}) has received a lot of attention in literature. 
The quantum chromodynamics (QCD) radiative corrections to it
were calculated in \cite{QCDrcor}, $\mathcal{O}(\alpha)$ EW corrections were 
calculated in \cite{EWrcor} and full $\mathcal{O}(\alpha)$ EW and 
$\mathcal{O}(\alpha_s)$ QCD corrections were studied in \cite{Belanger}.
Reaction (\ref{eetth}) was considered
in the kinematic region where the Higgs boson energy is close to its
maximal energy in \cite{Farrel}.
%and hence the next-to-leading-logarithmic corrections to
%the Higgs boson energy distribution 
%can be computed within the nonrelativistic effective theory.
Processes of the form $\epm \ra b\bar b b\bar b W^+W^- \ra b\bar b b\bar b
l^{\pm}\nu_lq\bar q'$ accounting for the signal of associated Higgs boson
and top quark pair production, as well as several irreducible background
reactions, were studied in \cite{Moretti} and EW contributions to
the leptonic and semileptonic reactions (\ref{ee8f}) have been 
computed in \cite{Schwinn}.
Pure off mass shell effects in 
$\epm \ra t^* \bar{t}^* H^* \ra b \bar b b \bar b u\bar d \mu^- \bar \nu_{\mu}$
have been discussed in \cite{offshell} and, for the on-shell Higgs boson,
in \cite{ustron07}. In the latter, also the off resonance background 
contributions have been calculated.
Moreover, feasibility of the measurement of the top--Higgs Yukawa coupling 
at the ILC in reaction (\ref{eetth}) was discussed in \cite{expfeas}.

\begin{figure}[htb]
\vspace{140pt}
\includegraphics{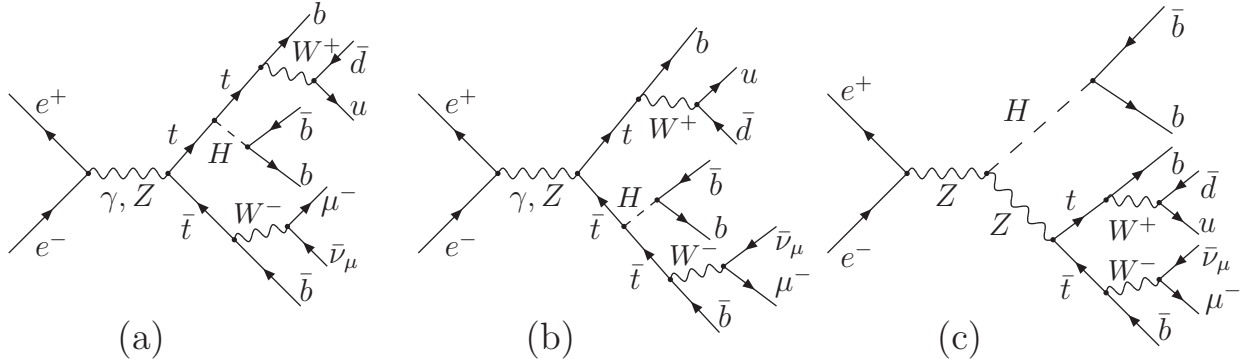}
\caption{Representative signal Feynman diagrams of reaction (\ref{udmn}).
The remaining diagrams are obtained by all possible permutations of 
the two $b$ and two $\bar b$ lines. The Higgs boson coupling to electrons
has been neglected.}
\label{fig:udmn}
\end{figure}

Already in the lowest order of SM reactions (\ref{ee8f})
receive contributions from tens thousands of Feynman diagrams, most of which 
comprise background to the resonant production and decay of the top quark pair
and Higgs boson. In order to show what a role the off resonance background effects
will play we select one reaction in each of the different detection channels of 
(\ref{ee8f}): the hadronic, semileptonic and leptonic channel
\bea
\label{udsc}
\epm &\ra& b \bar b b \bar b u\bar d s \bar c,\\
\label{udmn}
\epm &\ra& b \bar b b \bar b u\bar d \mu^- \bar \nu_{\mu},\\
\label{tnmn}
\epm &\ra& b \bar b b \bar b \tau^+ \nu_{\tau} \mu^- \bar \nu_{\mu},
\eea
respectively.
We will calculate the lowest order SM cross sections of 
(\ref{udsc}), (\ref{udmn}) and (\ref{tnmn}) 
with the complete set of the lowest order Feynman diagrams, both EW and QCD ones, 
and compare them with the corresponding signal cross sections of 
the associated resonant production and decay of the top quark pair
and Higgs boson, {\em i.e.} cross sections calculated with the signal
Feynman diagrams obtained from those depicted in Fig.~\ref{fig:eetth} by attaching
to each final state particle of (\ref{eetth}) lines representing its decay
products. Taking into account permutations of the $b$ and $\bar b$ quark lines
gives 20 signal diagrams for each of reactions (\ref{udsc}), (\ref{udmn}) and
(\ref{tnmn}). For illustration, we show representative
signal diagrams for (\ref{udmn}) in Fig.~\ref{fig:udmn}.
The comparison will be performed with different selections of cuts on particle
angles, energies, missing transverse energy and invariant masses which should allow
for correct identification of the signal and reduction of the background.

\section{Details of calculation}

Matrix elements of reactions (\ref{udsc}), (\ref{udmn}) and (\ref{tnmn}) 
have been generated automatically with a Fortran 90/95 program {\tt carlomat} 
\cite{carlomat} written by one of us (KK). 
Fermion masses, except for the neutrinos, can be kept nonzero in the program, 
but in order to speed up the calculation we neglect masses and Yukawa couplings
of the fermions lighter than $c$ quark and $\tau$ lepton. Then, 
taking into account both the EW and QCD lowest order contributions in the unitary
gauge, there are 39342, 26816 and 21214 Feynman diagrams for
(\ref{udsc}), (\ref{udmn}) and (\ref{tnmn}), respectively.
For each of the reactions {\tt carlomat} generates also dedicated phase space 
parametrizations which take into account mappings of peaks in the matrix 
element caused by propagators of massive unstable particles, of a photon, 
or a gluon in each Feynman diagram. This means that a number of different
phase space parametrizations generated is equal to a number of the Feynman 
diagrams. The phase space parametrizations 
are implemented into a multichannel Monte Carlo (MC) integration routine that 
performs integration over a 20-dimensional phase space.
However, only those parametrizations which result in different
phase space normalization are effectively used in the process of numerical 
integration.

Poles in the propagators of unstable particles
are regularized with constant particle widths
$\Gamma_a$  which are introduced through 
the complex mass parameters $M_a^2$ by making the substitution
\beq
\label{m2}
m_a^2 \;\ra \; M_a^2=m_a^2-im_a\Gamma_a, \qquad a=Z, W, H, t.
\eeq
Substitution (\ref{m2}) is made both in the $s$- and $t$-channel propagators.
The electroweak mixing parameter $\sin^2\theta_W$ can be kept real
\beq
\label{sw2}
\sin^2\theta_W=1-\frac{m_W^2}{m_Z^2},
\eeq
which is the approach usually referred to in the literature as the fixed width
scheme (FWS), or it may be defined as a complex quantity
\beq
\label{csw2}
\sin^2\theta_W=1-\frac{M_W^2}{M_Z^2},
\eeq
that is referred to as the complex mass scheme (CMS)
\cite{Racoon}. The latter has the advantage that it
preserves the lowest order Ward identities, which minimizes gauge invariance 
violation effects caused by (\ref{m2}).

The widths $\Gamma_a$, except for $\Gamma_Z$ whose actual value is rather
irrelevant in the context of associated top quark pair and Higgs boson
production and decay, are calculated in the lowest order of SM.
In the calculation of matrix elements with the helicity amplitude method,
use is made of the routines developed for a MC
program {\tt eett6f v.~1.0}, for calculating lowest order cross sections of 
$\epm \ra $ 6 fermions relevant for a $t\bar t$ pair production and decay
\cite{eett6f}, which have been tailored to meet needs of the automatic 
generation of amplitudes in {\tt carlomat}. MC summing over helicities is performed.

We have performed a number of tests of our results. First off all,
we have generated the matrix element of (\ref{udmn}) keeping the Yukawa
couplings and masses of all charged fermions, which results in 
56550 Feynman diagrams, and performed the phase space integration with
the corresponding number of different phase space parametrizations. The results
have agreed up to one standard deviation of the MC integration with 
those presented in the next
section, which have been obtained with 26816 Feynman diagrams after neglecting 
the light fermion masses and the Higgs boson
couplings to light fermions. Moreover, {\tt carlomat} offers an option
that allows to map poles caused by propagators
of internal particles which decay into 2, 3, 4, or more external particles.
This has led to different phase space parametrizations which all have given 
results consistent within 1--2 standard deviations.
In addition to these self-consistency checks we have compared our results for 
the cross sections of (\ref{udmn})
without QCD contributions with WHIZARD/OMEGA \cite{WHIZARD} obtaining agreement within 
one standard deviation. 

Generation of the Fortran code with {\tt carlomat} on 
a PC with the Pentium~4 3.0~GHz processor for any of reactions (\ref{udsc}), 
(\ref{udmn}), or (\ref{tnmn}) takes about 10 minutes CPU time. This relatively long time 
of the code generation is determined by a lot of write to and read from disk
commands which have to be introduced in order to circumvent limitations of 
the Fortran compilers concerning possible array sizes. The compilation time of generated
routines depends
strongly on a compiler used and an optimization option chosen. Typically,
for the considered reactions, it takes about one hour to compile all the
routines generated for each of the reactions. Most of the time is used for 
the compilation of the kinematical routines.
The execution time of the MC integration with about 2 million calls to the integrand
amounts typically to a few hours and, if the MC summing over polarizations is
employed, it is dominated by computation of the phase space normalization.

\section{Results}

The numerical results presented in this section have been obtained with the
following set of initial physical parameters: 
the Fermi coupling, fine structure constant in the Thomson limit and strong
coupling
\bea
\label{params1}
G_{\mu}=1.16639 \times 10^{-5}\;{\rm GeV}^{-2}, \qquad
\alpha_0=1/137.0359991, \qquad \alpha_s(m_Z)=0.1176,
\eea
the $W$- and $Z$-boson masses
\bea
\label{vmass}
m_W=80.419\; {\rm GeV},\qquad m_Z=91.1882\; {\rm GeV},
\eea
the top quark mass and the heavy external fermion masses of reactions (\ref{udsc}),
(\ref{udmn}) and (\ref{tnmn}) 
\beq
m_t=174.3\;{\rm GeV}, \quad m_b=4.8\;{\rm GeV}, \quad m_c=1.3\;{\rm GeV}, 
\quad m_{\tau}=1.77699\;{\rm GeV}. 
\eeq
%\beq
%\begin{array}{lll}
%m_u=5\;{\rm MeV}, & m_c=1.3\;{\rm GeV}, & m_t=174.3\;{\rm GeV}, \\
%m_d=10\;{\rm MeV},& m_s=200\;{\rm MeV}, & m_b=4.8\;{\rm GeV},\\
%m_e=me=0.51099892\;{\rm MeV}, & m_{\mu}=105.658369\;{\rm MeV}, &
%m_{\tau}=1.77699\;{\rm GeV}. 
%\end{array}
%\eeq
Light fermion masses are neglected except for in the test calculation of the 
cross section of (\ref{udmn}) with the full set of the lowest order Feynman diagrams,
mentioned in the previous section.
%that has been performed with
%\bea
%m_e=0.51099892\;{\rm MeV}, \quad  m_{\mu}=105.658369\;{\rm MeV}, \quad
%m_u=5\;{\rm MeV}, \quad m_d=10\;{\rm MeV}.
%\eea
The value of the Higgs boson mass is assumed at $m_H=130$~GeV.
Widths of unstable particles are calculated to the lowest order of SM resulting in
the following values:
\beq
\Gamma_t=1.53088\;{\rm GeV},\qquad \Gamma_W=2.04764\;{\rm GeV},\qquad
\Gamma_H=8.0540\;{\rm MeV}.
\eeq
The $Z$ boson width, whose actual value is not relevant in the calculation, is
put at its experimental value $\Gamma_Z=2.4952$~GeV. We use the CMS
with $\sin^2\theta_W$ given by (\ref{csw2}).

We identify jets with their original partons and define the following 
basic cuts which should allow to detect events with separate 
jets and/or isolated charged leptons:
\begin{itemize}
\item cuts on an angle between a quark and a beam, an angle between two quarks
and on a quark energy in reactions (\ref{udsc}), (\ref{udmn}) and (\ref{tnmn}): 
\beq
\label{cutsqq}
5^{\circ} < \theta (q,\mathrm{beam}) < 175^{\circ}, \qquad
\theta (q,q') > 10^{\circ}, \qquad E_{q} > 15\;{\rm GeV},
\eeq
\item cuts on angles between a charged lepton and a beam, a charged lepton 
and a quark and on energy of the charged lepton, $l=\mu,\tau$,
in reactions (\ref{udmn}) and (\ref{tnmn}): 
\beq
\label{cutslq}
5^{\circ} < \theta (l,\mathrm{beam}) < 175^{\circ}, \qquad
\theta (l,q) > 10^{\circ}, \qquad E_l > 15\;{\rm GeV}, 
\eeq
\item a cut on the missing transverse energy in reaction (\ref{udmn}),
\beq
\label{cutslqt}
/\!\!\!\!E^T > 15\;{\rm GeV},
\eeq
\item a cut on an angle between the two charged leptons and on the missing 
transverse energy in reaction (\ref{tnmn})
\beq
\label{cutsll}
\theta (l,l') > 10^{\circ}, \qquad /\!\!\!\!E^T > 30\;{\rm GeV}.
\eeq
\end{itemize}

The size of the background contributions is illustrated in 
Table~\ref{tab:angencuts}, where we give
results for the lowest order cross sections of reaction (\ref{udmn}) 
at a few centre of mass system (c.m.s.) energies in the presence
of basic cuts (\ref{cutsqq}), (\ref{cutslq}) and (\ref{cutslqt}).
The cross section $\sigma_{\rm all}$ calculated with the complete set of 
Feynman diagrams is shown in column 2. The cross section calculated without the gluon exchange
diagrams  $\sigma_{\rm no\;QCD}$ is given in column 3 and the signal cross 
section $\sigma_{\rm signal}$ calculated with
the 20 signal Feynman diagrams is presented in column 4. In order to see to
what extent the cuts reduce a signal of the associated production and decay 
of the top quark pair and Higgs boson we give the signal cross section
$\sigma_{\rm signal}^{\rm no\;cuts}$ and the signal 
cross section in the narrow width 
approximation (NWA) $\sigma_{\rm NWA}^{\rm no\;cuts}$ without cuts \cite{offshell}
in columns 5 and 6, respectively.
The numbers in parenthesis show the MC uncertainty of the last decimal.
We see that angular and energy cuts (\ref{cutsqq}), (\ref{cutslq}) 
and (\ref{cutslqt}) reduce
the signal by about 20\%, but they are not very efficient in reducing the background
contributions: both $\sigma_{\rm all}$ and $\sigma_{\rm no\;QCD}$ are substantially
larger than $\sigma_{\rm signal}$. In particular, at $\sqrt{s}=500$~GeV that is
the most realistic collision energy of ILC, the background
exceeds the signal by almost an order of magnitude. 
This is obviously caused by the fact that this collision energy is just above 
the $t\bar t H$ production threshold and the small phase space volume 
which is then available naturally reduces the signal cross 
section. Let us also note that, even though 
the cut on the energy of each quark has been imposed, the QCD background
contributions are quite sizeable, as can be seen by comparison of 
$\sigma_{\rm all}$ and $\sigma_{\rm no\;QCD}$.
\begin{table}
\begin{center}
\begin{tabular}{c|c|c|c|c|c}
\hline 
\hline 
\rule{0mm}{7mm} $\sqrt{s}$ [GeV]& $\sigma_{\rm all}$ [ab] 
& $\sigma_{\rm no\;QCD}$ [ab] & $\sigma_{\rm signal}$ [ab] 
& $\sigma_{\rm signal}^{\rm no\;cuts}$ [ab] 
             & $\sigma_{\rm NWA}^{\rm no\;cuts}$ [ab] \\[1.5mm]
\hline 
\rule{0mm}{7mm}  
  500 &  26.8(4) & 7.80(3) & 3.095(3) & 3.796(3) & 3.920(1)  \\ [1.5mm]
  800 & 100.2(8) & 66.8(1) & 46.27(2) & 58.36(2) & 60.03(2)  \\ [1.5mm]
 1000 &  93.1(3) & 61.4(1) & 40.18(2) & 51.74(2) & 52.42(3) \\ [1.5mm]
 2000 &  47.4(2) & 28.5(1) & 15.14(3) & 22.14(4) & 20.68(3) \\ [1.5mm]
\end{tabular} 
\end{center}
\caption{Cross sections of reaction (\ref{udmn}) at different c.m.s.
energies with cuts (\ref{cutsqq}), (\ref{cutslq}) 
and (\ref{cutslqt}) calculated: with the complete set of Feynman diagrams,
$\sigma_{\rm all}$, without gluon exchange diagrams, 
$\sigma_{\rm no\;QCD}$, and with only the signal diagrams of Fig.~\ref{fig:udmn},
$\sigma_{\rm signal}$. The last two columns show the total signal cross section 
$\sigma_{\rm signal}^{\rm no\;cuts}$ and the total cross section in NWA
$\sigma_{\rm NWA}^{\rm no\;cuts}$ of \cite{offshell} without cuts.
The numbers in parenthesis show the MC uncertainty of the last decimal.}
\label{tab:angencuts}
\end{table}

\begin{figure}[!tb]
\vspace{80pt}
\begin{center}
\setlength{\unitlength}{1mm}
\begin{picture}(35,35)(0,0)
\includegraphics{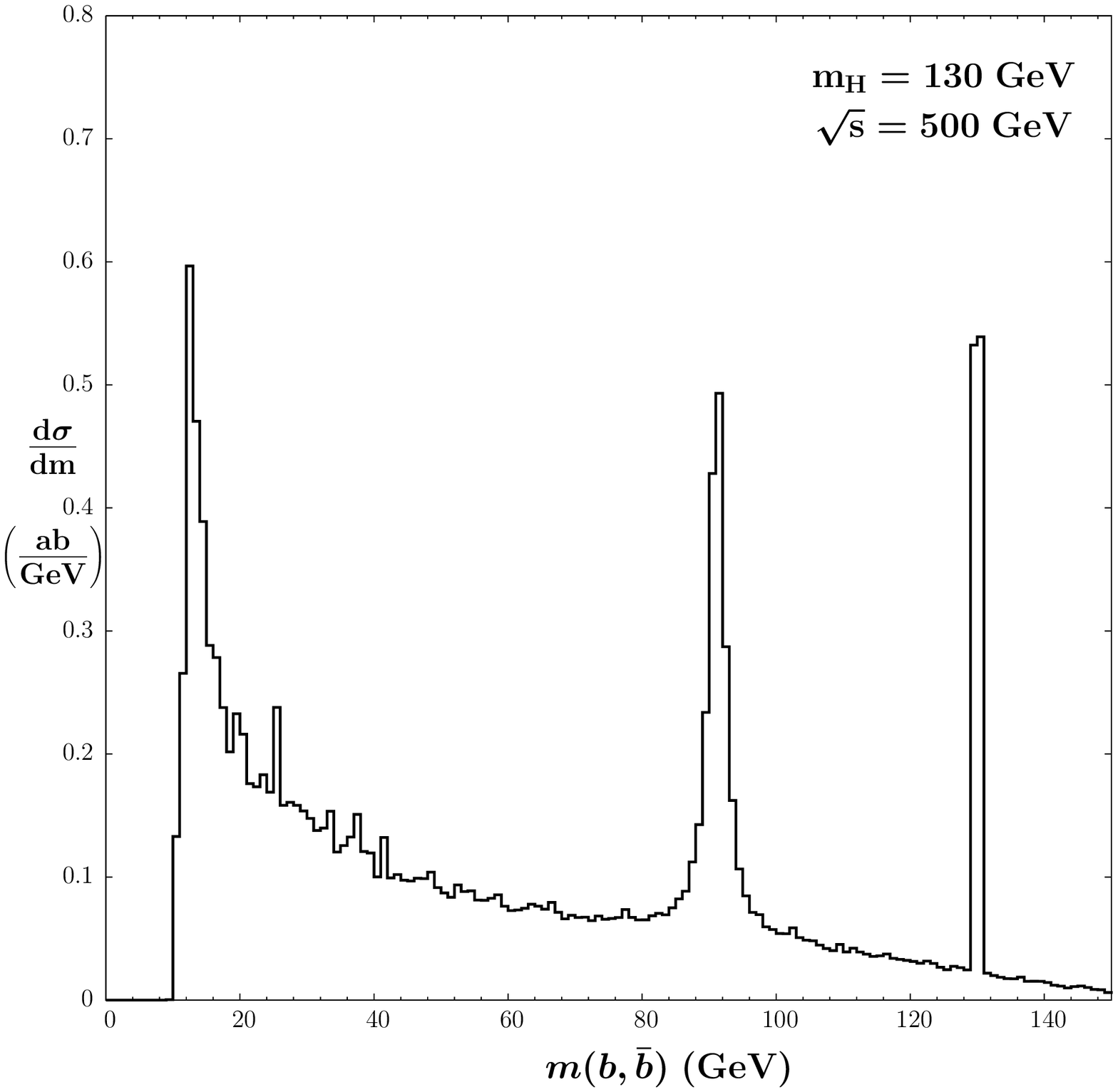}
\end{picture}
\hfill
\begin{picture}(35,35)(0,0)
\includegraphics{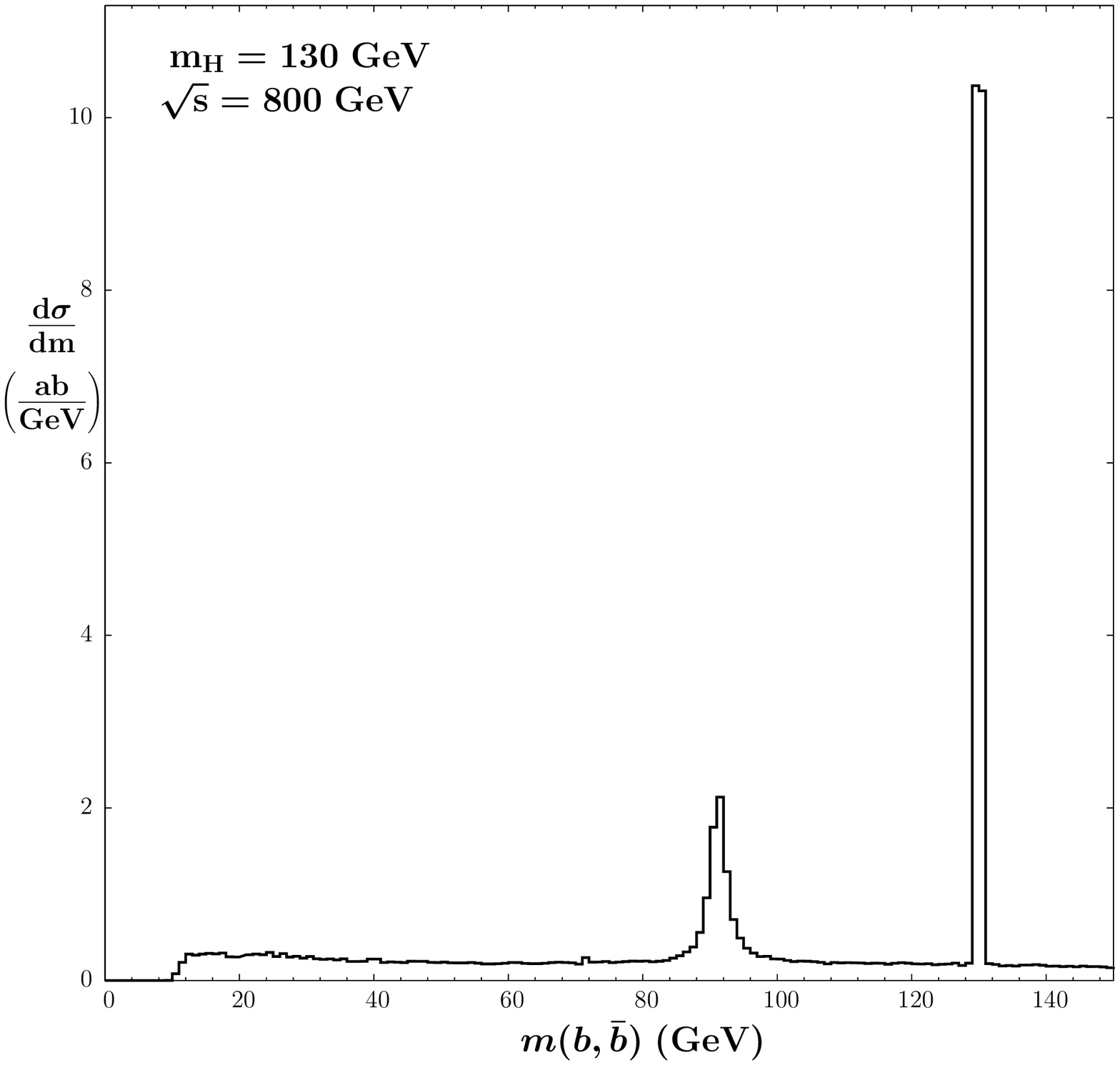}
\end{picture}
\end{center}
\vspace*{0.5cm}
\caption{Differential cross section of (\ref{udmn}) at $\sqrt{s}=500$~GeV (left)
$\sqrt{s}=800$~GeV (right) as a function of the invariant mass
of the two $b$ jets which remain after the other two $b$ jets have been associated
with top quarks: one  $b$ jet together with the two non $b$ jets have passed a cut
(\ref{cutmt}) and the $(b,\mu,/\!\!\!\!E_T)$ system has passed a cut (\ref{cutmtt}). 
The other cuts applied are given by 
(\ref{cutsqq}), (\ref{cutslq}), (\ref{cutslqt}), (\ref{cutmw}) and (\ref{cutmwt}).}
\label{fig:mbb}
\end{figure}

\begin{figure}[!tb]
\vspace{80pt}
\begin{center}
\setlength{\unitlength}{1mm}
\begin{picture}(35,35)(0,0)
\includegraphics{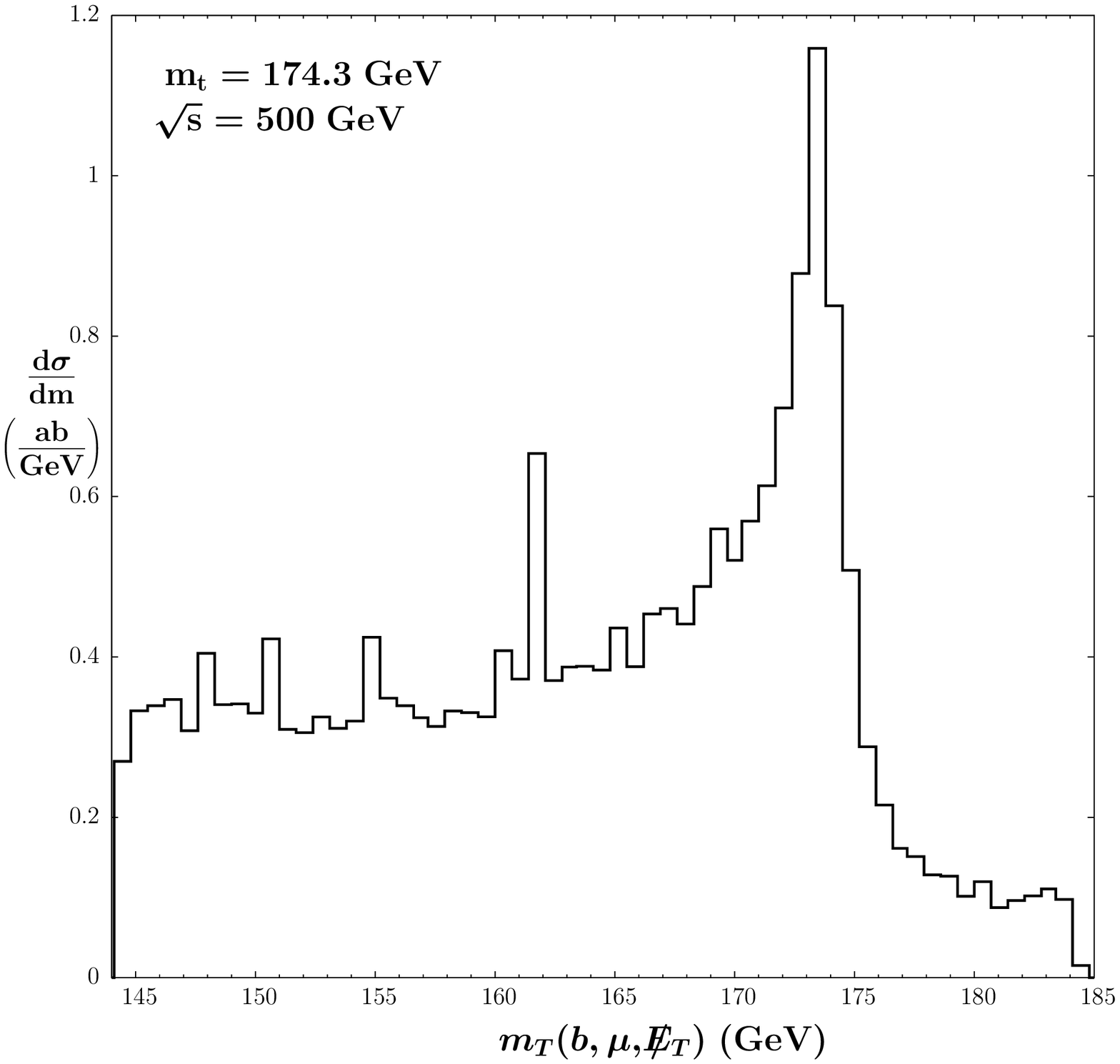}
\end{picture}
\hfill
\begin{picture}(35,35)(0,0)
\includegraphics{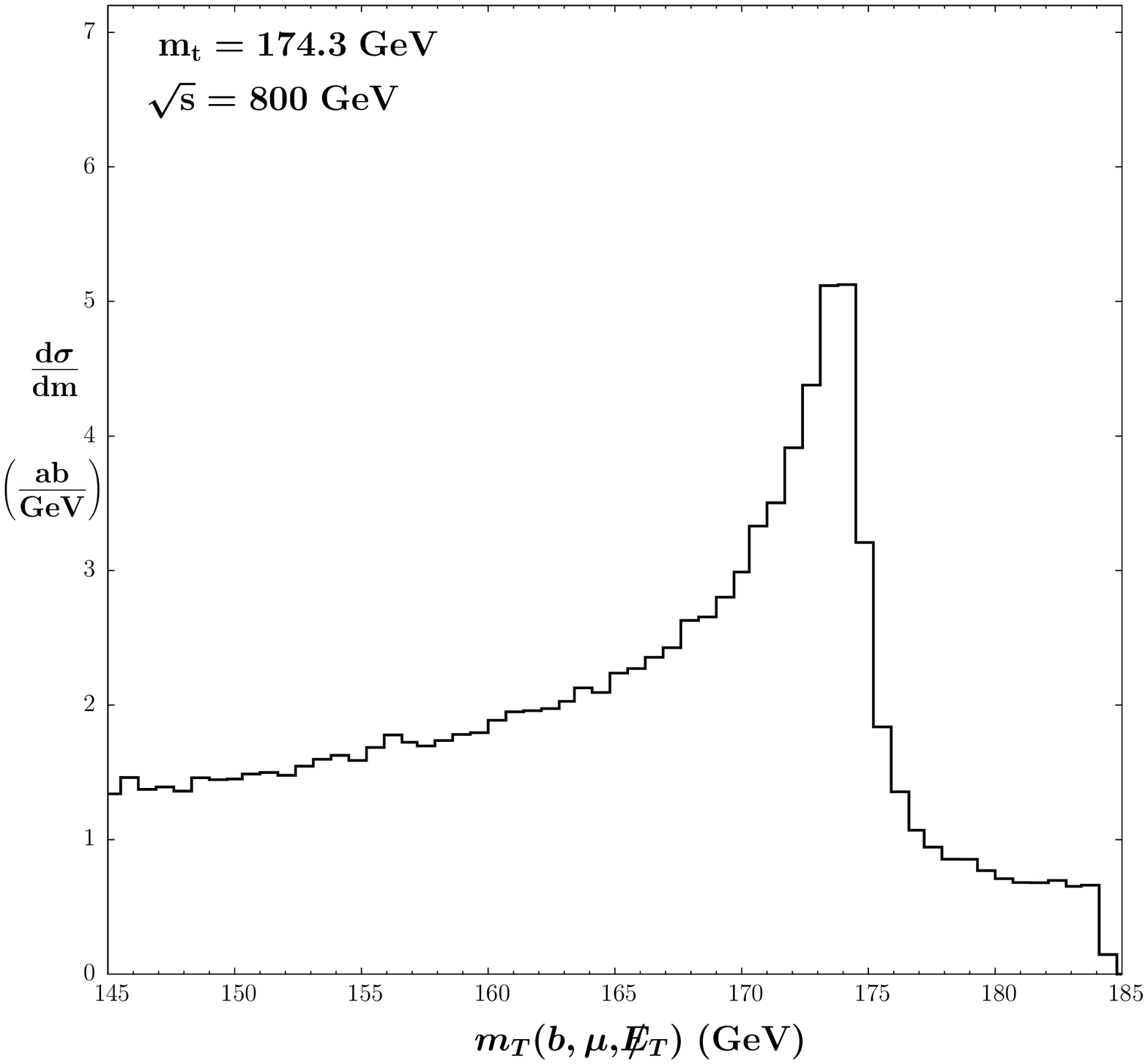}
\end{picture}
\end{center}
\vspace*{0.5cm}
\caption{Differential cross section of (\ref{udmn}) as a function of the transverse
mass, as given by (\ref{trmass}), of the $\left(b,\mu,/\!\!\!\!E_T\right)$ system that 
has passed a cut (\ref{cutmtt}) at $\sqrt{s}=500$~GeV (left) and
$\sqrt{s}=800$~GeV (right). The other cuts applied are given by 
(\ref{cutsqq}), (\ref{cutslq}), (\ref{cutslqt}), (\ref{cutmw}) and (\ref{cutmwt}).}
\label{fig:trmass}
\end{figure}

In order to reduce the background let us assume 100\% efficiency of $b$ tagging 
and define the following invariant 
mass cuts which should possibly allow to reconstruct $W$ bosons, $t$ quarks 
and the Higgs boson:
\begin{itemize}
\item a cut on the invariant mass of two non $b$ jets in reactions
(\ref{udsc}) and (\ref{udmn})
\beq
\label{cutmw}
60\;{\rm GeV} < \left[\left(p_{\sim b_1}+p_{\sim b_2}\right)^2\right]^{1/2} < 90
\;{\rm GeV}
\eeq
\item a cut on the transverse mass of the muon--neutrino system in reaction 
(\ref{udmn})
\bea
\label{cutmwt}
\left[m_{\mu}^2+2\left(m_{\mu}^2+\left|{\rm \bf p}_{\mu}^T\right|^2\right)^{1/2}
\left|\;/\!\!\!\!{\rm \bf p}^T\right|
-2{\rm \bf p}_{\mu}^T\cdot /\!\!\!\!{\rm \bf p}^T\right]^{1/2} < 90\;{\rm GeV},
\eea
\item a cut on the invariant mass of a $b$ jet, $b_1$, and two non 
$b$ jets, $b_{\sim b_1}, b_{\sim b_2}$, in reactions (\ref{udsc}) and (\ref{udmn})
\beq
\label{cutmt}
\left|\left[\left(p_{b_1}+p_{\sim b_1}+p_{\sim b_2}\right)^2\right]^{1/2} - m_t\right|
< 30\;{\rm GeV}
\eeq
\item a cut on the transverse mass $m_T$ of a $b$ quark, $b_2$, 
muon and missing transverse energy system in reaction (\ref{udmn})
\beq
\label{cutmtt}
m_t-30\;{\rm GeV} < m_T < m_t+10\;{\rm GeV},
\eeq
where
\bea
\label{trmass}
m_T^2=m^2+2\left(m^2+\left|{\rm \bf p}_{b_2}^T+{\rm \bf p}_{\mu}^T
\right|^2\right)^{1/2}/\!\!\!\!E^T
-2\left({\rm \bf p}_{b_2}^T+{\rm \bf p}_{\mu}^T\right)
\cdot /\!\!\!\!{\rm \bf p}^T, 
\eea
with $m$ being the invariant mass of the $b$-$\mu$ system given by
$m^2=\left(p_{b_2}+p_{\mu}\right)^2$.
\item an invariance mass cut on two $b$ jets, $b_3$ and $b_4$, 
in reactions (\ref{udsc}), (\ref{udmn}) and (\ref{tnmn})
\beq
\label{cutmh}
\left|\left[\left(p_{b_3}+p_{b_4}\right)^2\right]^{1/2} - m_H\right|
< m_{bb}^{\rm cut}, 
\eeq
with $m_{bb}^{\rm cut} = 20$~GeV, 5~GeV, or 1~GeV.
\end{itemize}
In order to justify the actual choice of invariant mass cuts in 
(\ref{cutmtt}) and (\ref{cutmh}) let us have a look at plots of some differential
cross sections of (\ref{udmn}).
In Fig.~\ref{fig:mbb}, we plot the differential cross section 
at $\sqrt{s}=500$~GeV (left) and $\sqrt{s}=800$~GeV (right) as a function 
of the invariant mass of the two $b$ jets which remain after the other two $b$ 
jets have passed cuts (\ref{cutmt}) and (\ref{cutmtt}) in the presence of 
angular and energy cuts (\ref{cutsqq}), (\ref{cutslq}) and (\ref{cutslqt}).
How well the top quark mass can be reconstructed from the
$\left(b,\mu,/\!\!\!\!E_T\right)$ system is illustrated
in Fig.~\ref{fig:trmass}, where we plot the 
differential cross section at $\sqrt{s}=500$~GeV (left)
and $\sqrt{s}=800$~GeV (right) as a function of 
transverse mass (\ref{trmass}) of the $\left(b,\mu,/\!\!\!\!E_T\right)$ system
that has passed a cut (\ref{cutmtt}).
The angular and energy cuts are given by (\ref{cutsqq}), (\ref{cutslq}) 
and (\ref{cutslqt}).

In Table~\ref{tab:mh20}, we show the lowest order $\sigma_{\rm all}$ and signal 
$\sigma_{\rm sig.}$ cross sections of reactions (\ref{udsc}), (\ref{udmn}) and (\ref{tnmn}) 
at a few c.m.s. energies in the presence of angular, energy and 
invariant mass cuts. 

\begin{table}
\begin{center}
\vspace*{-0.5cm}
\begin{tabular}{c|rr|rr|rr}
\hline 
\hline 
\rule{0mm}{6mm} $\sqrt{s}$ 
& \multicolumn{2}{c|}{$\epm \ra b \bar b b \bar b u\bar d s \bar c$}
& \multicolumn{2}{c|}{$\epm \ra b \bar b b \bar b u\bar d \mu^- \bar \nu_{\mu}$}
& \multicolumn{2}{c}{$\epm \ra b \bar b b \bar b \tau^+ \nu_{\tau}
   \mu^- \bar \nu_{\mu}$}\\[1.5mm]
\cline{2-7}
\rule{0mm}{6mm}  [GeV]& $\sigma_{\rm all}$ [ab] 
& $\sigma_{\rm sig.}$ [ab]  & $\sigma_{\rm all}$ [ab] 
& $\sigma_{\rm sig.}$ [ab]  & $\sigma_{\rm all}$ [ab] 
& $\sigma_{\rm sig.}$ [ab]  \\[1.5mm]
\hline 
\rule{0mm}{7mm}  
       & 13.88(6)  & 8.70(2)  & 3.50(2) & 2.384(3) &  4.03(9) & 0.863(2) \\ 
   500 & 10.17(4)  & 8.66(2)  & 2.62(1) & 2.332(3) &  1.89(7) & 0.864(2) \\ 
       &  9.07(4)  & 8.65(1)  & 2.37(1) & 2.312(3) &  1.09(2) & 0.860(1) \\[2.0mm]
       & 167.0(4)  & 128.4(1) & 43.6(1) & 33.93(2) & 23.28(5) & 13.48(1) \\ 
   800 & 139.1(3)  & 128.0(1) & 35.8(1) & 33.10(2) & 16.95(4) & 13.47(1) \\ 
       & 130.5(2)  & 127.7(1) & 33.4(1) & 32.82(2) & 14.44(4) & 13.46(1) \\[2.0mm]
       & 139.4(3)  & 109.1(1) & 35.3(1) & 27.94(1) & 21.70(5) & 12.09(1) \\ 
  1000 & 117.9(5)  & 109.0(1) & 29.5(1) & 27.52(1) & 15.41(3) & 12.10(1) \\ 
       & 110.6(2)  & 108.7(1) & 27.8(1) & 27.34(1) & 13.08(6) & 12.06(1) \\[2.0mm]
       & 44.5(2)   & 36.37(4) & 11.4(1) & 9.223(6) & 9.46(3)  & 4.95(2) \\
  2000 & 38.1(1)   & 36.23(4) & 9.76(4) & 9.157(6) & 6.49(3)  & 4.97(3) \\
       & 36.6(1)   & 36.09(2) & 9.25(4) & 9.136(6) & 5.42(2)& 4.97(1)
\end{tabular} 
\end{center}
\caption{Lowest order, $\sigma_{\rm all}$, and signal, $\sigma_{\rm sig.}$,
cross sections of reactions (\ref{udsc}), (\ref{udmn}) and (\ref{tnmn}). The first,
second and third row for each c.m.s. energy correspond to the Higgs boson invariant
mass cut $m_{bb}^{\rm cut}$ of (\ref{cutmh}) equal to, respectively, 20~GeV, 5~GeV and  1~GeV.
Other angular, energy and invariant mass cuts are specified for each reaction in the main text.
The numbers in parenthesis show the uncertainty of the last decimal.}
\label{tab:mh20}
\end{table}
The actual cutting procedures applied in different channels are the following.
\begin{itemize} 
\item In the hadronic channel, represented by reaction (\ref{udsc}), we first impose
angular and energy cuts (\ref{cutsqq}). Then we order the four non $b$ jets
in pairs and check if they satisfy $W$ boson identification criteria (\ref{cutmw}).
If so, we combine each pair with a $b$ jet and check if they obey
the $t$ identification criteria (\ref{cutmt}). Finally, we check if the remaining
two $b$ jets satisfy the Higgs boson identification cut (\ref{cutmh}).
\item In the semileptonic channel, represented by reaction (\ref{udmn}), we impose
angular and energy cuts (\ref{cutsqq}), (\ref{cutslq}) and the missing
energy cut (\ref{cutslqt}).
Then we check whether the two non $b$ jets satisfy a cut (\ref{cutmw}),
and if the muon and missing energy fulfil a transverse mass cut (\ref{cutmwt}).
If so, we combine a $b$ jet with the two non $b$ jets and check if they obey
the $t$ identification criteria (\ref{cutmt}). We select two of the remaining
three $b$ jets and check if they fulfil (\ref{cutmh}). Finally, we impose
a transverse mass cut (\ref{cutmtt}) on the other $b$ jet, the muon and missing
energy system.
\item In the leptonic channel, represented by reaction (\ref{tnmn}), we impose
angular and energy cuts (\ref{cutsqq}), (\ref{cutslq}) and (\ref{cutsll}).
We select two of the four $b$ jets and check if they satisfy the Higgs boson
identification criteria (\ref{cutmh}).
\end{itemize}
The cross sections in the first, second and third row for each c.m.s. energy correspond 
to the Higgs boson invariant mass cut $m_{bb}^{\rm cut}$ of (\ref{cutmh}) equal to, 
respectively, 20~GeV, 5~GeV and  1~GeV.
We see that the invariant mass cut on the two $b$ jets which reconstruct the Higgs boson
very efficiently reduces the background, practically for all the three detection channels of
(\ref{ee8f}) considered in Table~\ref{tab:mh20}. The smaller value of $m_{bb}^{\rm cut}$ 
the smaller is the background. The background is relatively the biggest in the leptonic channel, 
represented by reaction
(\ref{tnmn}), where we have not imposed invariant mass cuts which would allow for
reconstruction of $W$ bosons and $t$ quarks.

To which extent an extra cut on the energy of a $b$ quark
\beq
E_b > 40\;{\rm GeV}, \qquad {\rm or} \qquad E_b > 45\;{\rm GeV}.
\label{cuteb}
\eeq
can reduce the off resonance background contributions is shown in Table~\ref{tab:cuteb},
where we show the lowest order $\sigma_{\rm all}$ 
and signal $\sigma_{\rm sig.}$  cross section of (\ref{udmn}) and (\ref{tnmn}).
The first and
second row for each c.m.s. energy correspond to the Higgs boson invariant
mass cut $m_{bb}^{\rm cut}$ of (\ref{cutmh}) equal to, respectively, 20~GeV and 5~GeV.
We see that an extra cut (\ref{cuteb}) on energy of the $b$ quark efficiently reduces 
the background at $\sqrt{s}=500$~GeV, practically without altering the signal. 
However, it reduces both the background and the signal at higher c.m.s. energies.
For leptonic reaction (\ref{tnmn}), where no invariant mass cuts allowing either
$W$ bosons or $t$ quarks reconstruction have been imposed, the $b$ quark
energy cut (\ref{cuteb}) helps to reduce the background substantially.
\begin{table}
\begin{center}
\begin{tabular}{c|rr|rr|rr|rr}
\hline 
\hline 
\rule{0mm}{7mm} 
& \multicolumn{4}{c|}{$\epm \ra b \bar b b \bar b u\bar d \mu^- \bar \nu_{\mu}$}
& \multicolumn{4}{c}{$\epm \ra b \bar b b \bar b \tau^+ \nu_{\tau}
   \mu^- \bar \nu_{\mu}$}\\[1.5mm]
\cline{2-9}
\raisebox{2.0ex}[0pt]{$\sqrt{s}$}
& \multicolumn{2}{c|}{\rule{0mm}{7mm} $E_b > 40$~GeV} 
                                        & \multicolumn{2}{c|}{$E_b > 45$~GeV}
& \multicolumn{2}{c|}{$E_b > 40$~GeV} & \multicolumn{2}{c}{$E_b > 45$~GeV}\\[1.5mm]
\cline{2-9}
 \raisebox{2.0ex}[0pt]{[GeV]} & \rule{0mm}{7mm} $\sigma_{\rm all}$ [ab] 
& $\sigma_{\rm sig.}$ [ab]  & $\sigma_{\rm all}$ [ab] 
& $\sigma_{\rm sig.}$ [ab]  & $\sigma_{\rm all}$ [ab] 
& $\sigma_{\rm sig.}$ [ab]  & $\sigma_{\rm all}$ [ab] 
& $\sigma_{\rm sig.}$ [ab]  \\[1.5mm]
\hline 
\rule{0mm}{7mm}  
      & 3.25(1)  & 2.36(1) & 2.93(1) & 2.27(1) & 1.49(1)  & 0.850(2) 
& 1.22(1) & 0.810(2) \\ 
\raisebox{1.5ex}[0pt]{  500}
    & 2.57(1)  & 2.32(1) & 2.47(1) & 2.22(1) & 1.11(1) & 0.849(2) & 1.02(4) & 0.808(2) \\ [2.0mm]
      & 27.9(1)  & 23.72(3) & 24.0(1) & 20.58(3) & 13.58(4) & 9.38(1) 
& 11.51(2) & 8.14(1) \\ 
\raisebox{1.5ex}[0pt]{  800}
    & 24.4(1) & 23.07(3) & 21.1(1) & 20.03(3) & 10.99(2) & 9.37(1) & 9.44(2) & 8.13(1) \\ [2.0mm]
      & 22.7(1) & 19.50(2) & 20.1(1) & 17.49(2) & 12.84(3) & 8.48(1) 
& 11.18(2) & 7.59(1) \\ 
\raisebox{1.5ex}[0pt]{ 1000}
    & 20.0(1) & 19.20(2) & 18.0(1) & 17.21(2) & 10.07(2) & 8.47(1) & 8.92(2) & 7.59(1) \\ [2.0mm]
      & 8.59(4)  & 7.35(1)  & 8.10(3) & 6.95(1)  & 6.45(2)  & 4.02(2)
& 5.93(2) & 3.78(2) \\ 
\raisebox{1.5ex}[0pt]{ 2000}
      & 7.61(3) & 7.28(1) & 7.27(6) & 6.90(1) & 4.84(2) & 4.01(2) & 4.51(1) & 3.84(2)
\end{tabular} 
\end{center}
\caption{Lowest order, $\sigma_{\rm all}$, and signal, $\sigma_{\rm sig.}$,
cross sections of reactions (\ref{udmn}) and (\ref{tnmn})
with an extra cut (\ref{cuteb}) on the $b$ quark energy. The first and
second row for each c.m.s. energy correspond to the Higgs boson invariant
mass cut $m_{bb}^{\rm cut}$ of (\ref{cutmh}) equal to, respectively, 20~GeV and 5~GeV.
Other angular, energy and invariant mass cuts are specified for each reaction in the main text.
The numbers in parenthesis show the uncertainty of the last decimal.}
\label{tab:cuteb}
\end{table}

\section{Conclusions}

We have calculated the lowest order SM cross sections of reactions
(\ref{udsc}), (\ref{udmn}) and (\ref{tnmn}) representing the hadronic, semileptonic
and leptonic detection channels of the associated production 
of the top quark pair and Higgs boson at the ILC.
In the calculation, we have taken into account complete sets of the lowest 
order Feynman diagrams, both EW and QCD ones. A comparison of the cross sections
with the corresponding signal cross sections of the associated production 
and decay of the top quark pair and Higgs boson in the presence of angular
and energy cuts has shown that the off resonance
background is large. Imposing invariant mass cuts, which allow for reconstruction
of $W$ bosons, $t$ quarks and the Higgs boson, reduces the background. 
In particular, a cut (\ref{cutmh}) on the invariant mass of the two $b$ quark system
that reconstructs the Higgs boson is very efficient: the smaller the cut
the better is the background reduction.
A cut on the energy of the slowest $b$ jet can further reduce the background
at $\sqrt{s}=500$~GeV, but for higher c.m.s. energies it becomes  less efficient, as
it reduces the signal as well. 

Acknowledgements: This work is supported in part by the Polish Ministry of Science 
under Grant No. N N519 404034
and by European Community's Marie-Curie Research Training Network
under contracts MRTN-CT-2006-035482 (FLAVIAnet) and MRTN-CT-2006-035505 (HEPTOOLS).


\begin{thebibliography}{99}
\bibitem{ILC} James Brau, Yasuhiro Okada, Nicholas Walker, {\it et al.}
              [ILC Reference Design Report Volume 1 - Executive Summary],
              arXiv:0712.1950;\\
              J.A. Aguilar-Saavedra {\it et al.} [ECFA/DESY LC Physics
              Working Group Collaboration], arXiv:hep-ph/0106315;\\
              T.~Abe {\it et al.}, [American Linear Collider Working Group
              Collaboration],
%              SLAC-R-570, {\it Resource book for Snowmass 2001},
               arXiv:hep-ex/0106056;\\
               K.~Abe {\it et al.} [ACFA Linear Collider Working Group
               Collaboration], arXiv:hep-ph/0109166.
\bibitem{eetth} A. Djouadi, J. Kalinowski, P.M. Zerwas, 
                Mod. Phys. Lett. {\bf A7} (1992) 1765;\\
                A. Djouadi, J. Kalinowski, P.M. Zerwas, Z. Phys 
                {\bf C54} (1992) 255.
\bibitem{LEPdir} R. Barate {\em et al.}, Phys. Lett. {\bf B565} (2003) 61.
                   % arXiv:hep-ex/0306033
\bibitem{Hmass} The LEP Collaborations and the LEP electroweak
                working group, arXiv:hep-ex/0612034v2, and references therein;\\
                Ch. Parkes, International Conference on High Energy Physics,
                {\tt http://lephiggs.web.cern.ch/}, July 2006;\\
                B. Kilminster, arXiv:hep-ex/0611001, to appear in the 
                proceedings of 33rd International Conference on High 
                Energy Physics (ICHEP 06), Moscow, Russia, 
                26 Jul. -- 2 Aug. 2006.
\bibitem{QCDrcor} S. Dittmaier, M. Kramer, Y. Liao, M. Spira, P.M. Zerwas, 
                  Phys. Lett. {\bf B441} (1998) 383;\\
                  S. Dittmaier, M. Kramer, Y. Liao, M. Spira, P.M. Zerwas, 
                  Phys. Lett. {\bf B478} (2000) 247;\\
        %hep-ph/9808433 Higgs radiation off top quarks in e+ e- collisions.
                  S. Dawson, L. Reina, Phys Rev. {\bf D57} (1998) 5851;\\
                  S. Dawson, L. Reina, Phys Rev. {\bf D59} (1999) 054012.
        %hep-ph/9808443 QCD corrections to associated Higgs boson heavy 
                  %quark production.
\bibitem{EWrcor} Yu You {\it et al.}, Phys. Lett. {\bf B571} (2003) 85;\\
                 %hep-ph/0306036
\bibitem{Belanger} G. B\'elanger {\it et al.}, Phys. Lett. {\bf B571} (2003) 
                   163;\\
                   %hep-ph/0307029;  Full O(alpha) electroweak and O(alpha(s)) 
                   % corrections to e+ e- ---> t anti-t H.
                   A. Denner, S. Dittmaier, M. Roth, M.M. Weber, Phys. Lett. 
                   {\bf B575} (2003) 290; \\ %hep-ph/0307193
                   A. Denner, S. Dittmaier, M. Roth, M.M. Weber, Nucl. Phys.
                   {\bf B680} (2004) 85. %hep-ph/0309274
\bibitem{Farrel} C. Farrell, A.H. Hoang, Phys. Rev. {\bf D72} (2005) 014007;\\
                 C. Farrell, A.H. Hoang, Phys. Rev. {\bf D74} (2006) 014008.
\bibitem{Moretti} S. Moretti, Phys. Lett. {\bf B452} (1999) 338. %hep-ph/9902214
\bibitem{Schwinn} C. Schwinn, arXiv:hep-ph/0412028.
\bibitem{offshell} K. Ko\l odziej, S. Szczypi\'nski, Acta Phys. Pol. {\bf B38} 
                   (2007) 2565--2576.%, hep-ph/0612183.
\bibitem{ustron07} K. Ko\l odziej, S. Szczypi\'nski, 
                   Acta Phys. Pol. {\bf B38} (2007) 3609.
\bibitem{expfeas} H. Baer, S. Dawson, L. Reina, Phys Rev. {\bf D61} (2000) 
                  013002;\\
                  A. Juste, G. Merino, arXiv:hep-ph/9910301;\\
                  A. Juste, ECONF C0508141:ALCPG0426, 2005, 
                  arXiv:hep-ph/0512246;\\
                  A. Gay, arXiv:hep-ph/0604034.
\bibitem{carlomat} K.~Ko\l odziej, {\em ``{\tt carlomat}, a program for 
                   automatic computation of multiparticle cross sections''}, 
                   in preparation.
\bibitem{Racoon} A. Denner, S. Dittmaier, M. Roth, D. Wackeroth,
                 Nucl. Phys. {B560} (1999) 33 and
                 Comput. Phys. Commun. {153} (2003) 462.
\bibitem{eett6f} K. Ko\l odziej, Comp. Phys. Commun. {\bf 151} (2003) 339.
\bibitem{WHIZARD} M. Moretti, T. Ohl, J. Reuter, IKDA 2001/06-rev, 
                  LC-TOOL-2001-040-rev, hep-ph/0102195-rev;\\
                  W.~Kilian, T.~Ohl, J.~Reuter,
                  %``WHIZARD: Simulating Multi-Particle Processes at LHC and ILC,''
                  arXiv:0708.4233.
\end{thebibliography}
\end{document}